\documentclass[prr,twocolumn,longbibliography]{revtex4-2}
\usepackage{graphicx} % Required for inserting images
\usepackage{amsmath}
\usepackage{hyperref}
\usepackage{graphicx}
\usepackage{color}
\newcommand{\bk}{{\bf k}}

\newcommand{\br}{{\bf r}}

\newcommand{\LSCO}{La$_{2-x}$Sr$_x$CuO$_4$}
\newcommand{\BSCCO}{Bi$_2$Sr$_2$CaCu$_2$O$_{8+\delta}$}
\newcommand{\BSCO}{Bi$_2$Sr$_2$CuO$_{6+\delta}$}

\begin{document}
\title{Inhomogeneity, Fluctuations, and Gap Filling in Disordered Overdoped Cuprates}
\author{Miguel Antonio Sulangi$^1$, Willem Farmilo$^2$, Andreas Kreisel$^3$, Mainak Pal$^4$, W. A. Atkinson$^2$, P. J. Hirschfeld$^4$ }
\affiliation{{}$^1$National Institute of Physics, University of the Philippines Diliman, Quezon City 1101, Philippines \\
    {}$^2$Department of Physics and Astronomy, Trent University, Peterborough, Ontario K9L 0G2, Canada \\
 {}$^3$Niels Bohr Institute, University of Copenhagen, DK-2200 Copenhagen, Denmark \\
  {}$^4$Department of Physics, University of Florida, Gainesville, FL 32611, USA
}

\date{\today}
\begin{abstract}
Several recent experiments have challenged the premise that cuprate high-temperature superconductors approach conventional Landau-BCS behavior in the high-doping limit.  We argue, based on an analysis of their superconducting spectra, that  anomalous properties seen in the most-studied overdoped cuprates require a  pairing interaction that is strongly inhomogeneous on nm length scales. This is consistent with recent proposals that the ``strange-metal'' phase above $T_c$ in the same doping range arises from a spatially random interaction.  We show, via  mean-field Bogoliubov-de Gennes (BdG) calculations and time-dependent Ginzburg-Landau (TDGL) simulations, that key features of the observed tunneling spectra are reproduced when both inhomogeneity and thermal phase fluctuations are accounted for.  In accord with experiments, BdG calculations find that low-$T$ spectra are highly inhomogeneous and exhibit a low-energy spectral shoulder and broad coherence peaks.  However, the spectral gap in this approach becomes homogeneous at high $T$, in contrast to experiments.  This  is resolved when thermal fluctuations are included; in this case, global phase coherence is lost at the superconducting $T_c$ via a broadened BKT transition, while robust phase-coherent superconducting islands persist well above $T_c$.  The local spectrum remains inhomogeneous at $T_c$, and the gap is found to fill instead of close with increasing temperature.  

\end{abstract}
\maketitle

\section{Introduction}
Underdoped and optimally doped cuprates represent  the epitome of doped Mott physics  due to their proximity to the parent insulator, and exhibit not only superconductivity but other  intertwined phases as well.   Until relatively recently, however, it was generally accepted that the {\it overdoped} materials are uncomplicated by a pseudogap or other competing phases
and should neatly approach the Landau-BCS paradigm with increasing doping.

Numerous features of the overdoped regime challenge this notion.  
First, the superfluid density at zero temperature in high quality \LSCO (LSCO) films was found to be proportional to the critical temperature $T_c$\;\cite{Bozovic:2016Dependence,Mahmood:2019microwave}, in stark contrast to the BCS prediction that it should be equal to the electron density and independent of $T_c$.  Second, scanning tunneling spectroscopy (STS) experiments in Pb- and La-doped \BSCO (Bi-2201) repeatedly observe that the spectrum has a two-gap structure, with a homogeneous low-energy gap that is ascribed to superconductivity and an inhomogeneous high-energy gap that is reminiscent of the pseudogap found in underdoped cuprates\;\cite{Boyer:2007STM,He:2014STM,Li:2021CDW,Ye:2024Emergent}.  A  two-gap structure has also been observed in overdoped \BSCCO{} (Bi-2212)\;\cite{Alldredge:2008evolution,Pushp:2009STM}, though the two gaps are close in energy and hard to resolve.  Third, overdoped Bi-2201 and Bi-2212 are highly inhomogeneous, with a spectral gap that varies by a factor of two or more over distances of a few nm\;\cite{Pasupathy:2008STM,Parker:2010,Boyer:2007STM,Li:2021CDW,Tromp:2023Puddle,Ye:2024Emergent}.  Temperature-dependent STS\;\cite{Gomes2007,Pasupathy:2008STM,Li:2022} and angle-resolved photoemission spectroscopy (ARPES) experiments\;\cite{hashimoto2014energy,He:2021ARPES,Chen2022} have shown that in both materials these spectral gaps close well above $T_c$, raising  questions about the nature of the superconducting state.  Magnetoresistance measurements on {LSCO} found evidence for superconducting fluctuations extending a few tens of kelvin above $T_c$\;\cite{Rourke:2011phase} throughout the overdoped regime; from this, it was inferred that $T_c$ occurs at a Berezinskii-Kosterlitz-Thouless (BKT) transition.  However, magnetic susceptibility measurements\;\cite{Sonier:2008,Koike:2008inhomogeneous,Li:2021Granular} in LSCO find diamagnetic signatures extending up to much higher temperatures.  These invite a ``granular'' interpretation, in which superconducting $T_c$ is determined by the strength of Josephson coupling between robustly superconducting islands\;\cite{Spivak2009}.

Furthermore, it has been argued that granularity above $T_c$ arises naturally if the overdoped cuprates contain a sufficient density of pair-breaking impurities\;\cite{Li:2021Granular} so that superconductivity persists only in rare regions.  However, numerical simulations cast doubt that heavily disordered $d$-wave superconductors are truly granular\;\cite{Breio2022}.  Indeed, the disorder levels suggested in Ref.\;\cite{Li:2021Granular} would generate a much larger residual density of states\;\cite{Pal:2023} than is actually observed by ARPES measurements on very overdoped Bi-2212\;\cite{Yin:2024ARPES}, and suppress the average $T_c$ to zero.
\begin{figure*}[tb]
\includegraphics[width=\textwidth]{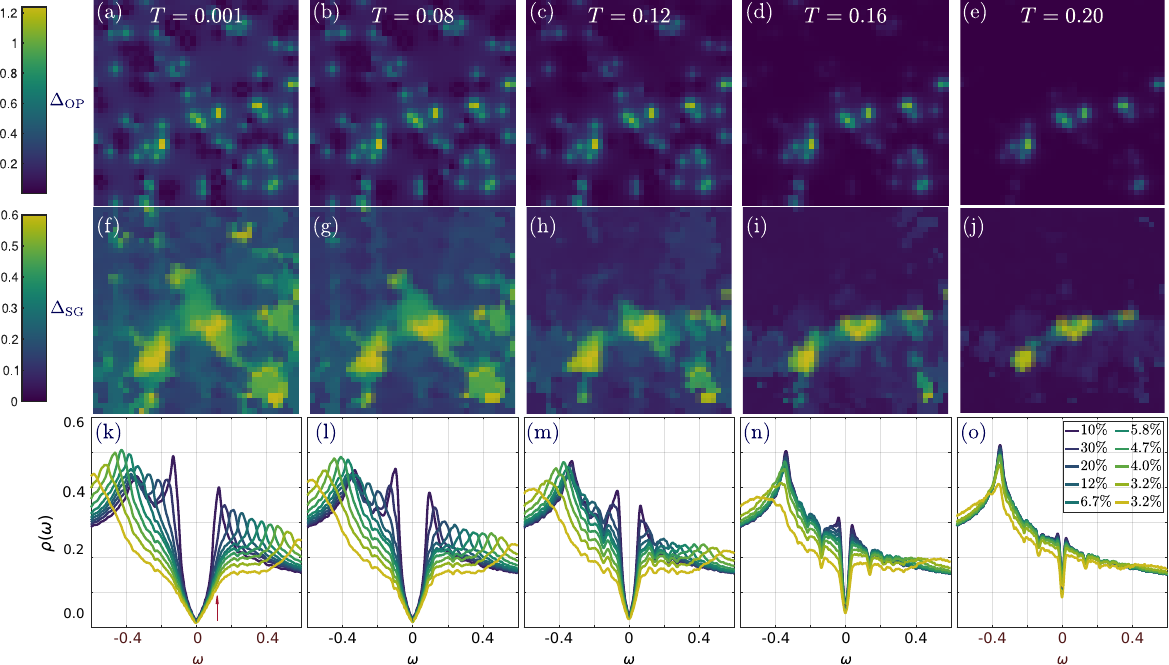}
	\caption{Spatial maps of the $d$-wave order parameter $\Delta_{OP}$ (top row) and spectral gap $\Delta_{SG}$ (middle row) for a single realization of the pairing interaction.
    The LDOS (bottom row) is calculated for an ensemble of 50 random pairing realizations and for $15\times 15$ supercells.  Sites are binned by the value of $\Delta_{SG}$ at $T=0.001$, and the average LDOS is shown for each bin (bottom row).  The same sites are in each bin at all temperatures. Ten equal-width bins spanning $0.085 < \Delta_{SG} < 0.605$ are used; the legend shows the fraction of sites in each bin and darker colors correspond to lower-energy bins.
    Results are shown at five temperatures ranging from $T\ll T_\mathrm{mf}$ to $T \approx T_\mathrm{mf}$,   for an impurity concentration of 7.5\%. The red arrow in (k) indicates the subgap shoulder.}
	\label{fig:op_sg_ldos_yukawa_c}
\end{figure*}

Here we propose a rather different picture.  We suggest that the observed spectroscopic properties of overdoped Bi-2201 and Bi-2212 depend on two ingredients: a localized pairing interaction whose strength varies by an order of magnitude  on a scale of nanometers, and \textit{inhomogeneous} phase fluctuations. We further argue that pair-breaking disorder, which is essential to explain anomalous low-energy properties like the superfluid density\;\cite{Lee-Hone:2017disorder,Lee-Hone:2018optical,Lee-Hone:2020,Ozdemir:2022}, is less relevant on the energy scale of the gap.   Our calculations show that a two-gap spectrum with a homogeneous subgap and rapidly varying large gap emerges naturally when the inhomogeneity in the pair potential is sufficiently strong, without the need to appeal to a pseudogap, and that both gaps are associated with superconductivity.  Furthermore, thermal fluctuations have a dramatic effect on the subgap, which fills with increasing $T$ and disappears when superconducting phase coherence is lost.  We find that both Bi-2201 and Bi-2212 can be understood within our model, with the key difference between the two systems
being the degree of inhomogeneity.  Our calculations suggest that $T_c$ in these cuprates is determined by a 
BKT-like transition that is broadened by disorder\;\cite{Benfatto:2009,Mahmood:2022} and interlayer coupling\;\cite{Benfatto:2007kosterlitz} rather than a granular transition, and that a weak diamagnetic signal persists above $T_c$ because of isolated superconducting islands with anomalously large pairing interactions.

The idea that phase-incoherent pairing can open a partial gap without superconductivity was originally explored in the context of underdoped cuprates\;\cite{franz1998phase,kwon1999effect}.  It was deemed an unlikely model for the pseudogap because of the unreasonably large temperature window ($\sim 100$~K) over which a vortex-antivortex plasma must persist\;\cite{franz1998phase}.  Nonetheless, such models
capture coarse features of ARPES and scanning tunneling microscopy (STM) data   near $T_c$\;\cite{Eckl:2002pair,mayr2005phase,valdez2006single,Chubukov:2007,Berg:2007evolution,valdez2008effects,alvarez2008fermi,Atkinson:2012,Hayward:2014angular}. At low $T$, Nunner \textit{et al.}\;\cite{Nunner:2005dopant} and Fang \textit{et al.}\;\cite{Fang2006} showed that pairing inhomogeneity is responsible for two key properties of  spectra in underdoped Bi-2212\;\cite{McElroy:2005}---that they are largely particle-hole symmetric and homogeneous at low energies, and that there is an inverse correlation between coherence peak energy and peak height---that cannot be explained by conventional impurity models.  The main feature of the current work is the focus on how large pairing  and phase fluctuations modify local spectra in a strongly inhomogeneous superconductor near the transition.

\section{Results: mean-field theory}
We first solved the
Bogoliubov-de Gennes (BdG) equations on a $40\times 40$ lattice for the tight-binding Hamiltonian
\begin{equation}
    H=\sum_{ij\sigma} t_{ij}c^\dagger_{i\sigma}c_{j\sigma} + \sum_{ij} \left(\Delta_{ij}^\ast c_{j\downarrow} c_{i\uparrow} + \Delta_{ij} c^\dagger_{i\uparrow} c^\dagger_{j\downarrow}\right ),
\end{equation}
with $t_{ij} = t$ ($t_{ij} = -0.35t$) for nearest-neighbor (next-nearest-neighbor) lattice sites $i$ and $j$, and with $\Delta_{ij} = V_{ij}\langle c_{j\downarrow}c_{i\uparrow}\rangle$ determined self-consistently along each bond. $V_{ij}$ is nonzero for nearest-neighbor sites only (see Appendix), and $\Delta_{ij}$ is constrained to be real so that there are no spontaneous supercurrents. We set $t=-1$ so that $|t|$ is the unit of energy, and the chemical potential is adjusted to maintain an electron density $n=0.85$ per unit cell.  Note that the value of $n$ does not matter for this work so long as the Fermi surface is qualitatively reasonable.
As in Ref.~\cite{Nunner:2005dopant}, the model assumes that the pairing interaction $V_{ij}$ has a weak homogeneous component $V_0$ that is modulated locally near randomly-distributed impurities with a Yukawa spatial profile, $V_I e^{-r/\lambda}/r$, with $r$ the distance to the impurity and $\lambda$ the range of the modulation (see Sec.~\ref{sec_inhomogeneity} for details).  Importantly, $\lambda$ is comparable to the typical superconducting coherence length $\xi$ at $T=0$, so that there are large nanoscale spatial fluctuations of the order parameter.

Figure~\ref{fig:op_sg_ldos_yukawa_c} compares the $d$-wave component of the order parameter, spectral gap, and bin-averaged local density of states (LDOS) for temperatures ranging from near zero to near the mean-field transition temperature, $T_\mathrm{mf}$.  The order parameter at site $i$ is $\Delta_{OP,i}=|\sum_\delta (-1)^{\delta_y}\Delta_{i,i+\delta}|$, and
the spectral gap $\Delta_{SG,i}$ is the energy of the largest positive-energy peak in the LDOS at site $i$. The homogeneous interaction ($V_0 = 0.8$) by itself gives $\Delta_{OP} = 0.19$ at $T=0.001$ and has a critical temperature $T_c = 0.085$.

\begin{figure}[tb]
	\centering
    \includegraphics[width=1.00\columnwidth]{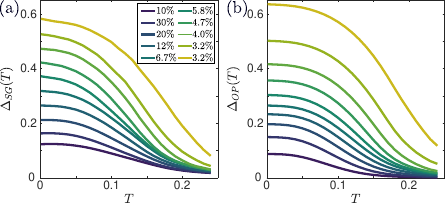}
    \caption{The binned spectral gap (a) and order parameter (b) as functions of temperature. At the lowest measured temperature, each site is sorted its local value of $\Delta_{SG}$ and put into one of ten bins; the values of $\Delta_{SG}$ and $\Delta_{OP}$ are then averaged over each bin at each temperature.}
	\label{fig:sgbinned}
\end{figure}

Approximately 40\% of the lattice sites belong to the 1st or 2nd bins, see legend in Fig.~\ref{fig:op_sg_ldos_yukawa_c}(o).  At low $T$, these ``small-gap'' sites are characterized by a canonical $d$-wave density of states with well-defined coherence peaks. Spectra associated with ``large-gap'' sites, belonging to the 3rd or higher bin, deviate from a simple $d$-wave form:  they have a two-gap structure, with low-energy shoulders and high-energy coherence peaks. The shoulders are often interpreted as a subgap, and we adopt that terminology.
The shoulders occur at the same energies for all bins, and coincide with the coherence peak energy of the 2nd bin. We ascribe the shoulders to proximity coupling to small-gap sites; that is, they are the extension  of neighboring coherence peaks onto large-gap sites\;\cite{Fang2006}.
Where they are distinct, the spectral gap (which coincides with the large gap) reflects the local order parameter, and the subgap reflects the typical order parameter.  

The correspondence between $\Delta_{SG}$ and $\Delta_{OP}$ is not perfect \cite{Sulangi:2021correlations}: spatial variations of $\Delta_{SG}$ are a factor $\sim 2$ less than $\Delta_{OP}$. Nonetheless, $\Delta_{SG}$ appears granular at high $T$, with robust superconducting islands embedded in a weakly superconducting, nearly uniform background.  The granularity is not reflected in the subgap, which is spatially uniform.  Indeed, it is striking that an order parameter with spatial variations exceeding an order of magnitude can generate an essentially homogeneous low-energy LDOS.   Equally remarkable, the subgap persists to temperatures well above the subgap energy, where superconductivity would be suppressed in a homogeneous BCS superconductor.
Its persistence at these elevated temperatures is the result of proximity coupling to nearby regions with large pairing interactions. 

Because the typical coherence length is larger than the interimpurity spacing and grows near $T_c$, this proximity coupling produces a spectrum that is surprisingly homogeneous and $d$-wave-like near $T_\mathrm{mf}$.   
{This is evident in the binned LDOS shown in Fig.~\ref{fig:op_sg_ldos_yukawa_c}, and is summarized in Fig.~\ref{fig:sgbinned}, which shows the $T$-dependence of $\Delta_{OP}$ and $\Delta_{SG}$,
both sorted into ten equally sized bins according to the local values of $\Delta_{SG}$.
Both are inhomogeneous at low $T$; however, by $T=0.2 \approx T_\mathrm{mf}$, $\Delta_{SG}$ is nearly the same for all bins, apart from the largest-gap bin, while $\Delta_{OP}$ varies by an order of magnitude or more between small- and large-gap bins. 
The homogeneity of $\Delta_{SG}$ near $T_\mathrm{mf}$ is incompatible with temperature-dependent STS experiments, primarily on Bi-2212\;\cite{Parker:2010,Pasupathy:2008STM}, which show that the spectral gap is strongly inhomogeneous near $T_c$, and that it fills rather than closes.

\section{Results: fluctuations}
To address these discrepancies, we consider order-parameter fluctuations in tandem with inhomogeneity. We first solved a set of time-dependent Ginzburg-Landau (TDGL) equations with thermal noise to obtain an ensemble of order parameter ``snapshots.''  These calculations naturally allow  both amplitude and complex phase fluctuations.  Inhomogeneity is built into the model via the lowest-order Ginzburg-Landau (GL) parameter, which has a similar Yukawa spatial dependence as the BdG pairing interaction (see section \ref{sec_inhomogeneity}).  Once order-parameter ensembles were generated, we calculated the LDOS at selected sites using a microscopic non-self-consistent BdG Hamiltonian, with the same tight-binding parameters as above and taking the TDGL snapshots as input.  TDGL calculations were for a single disorder realization on a $300\times 300$ lattice.

This approach  follows a broad body of work in which an ansatz is made to generate an ensemble of order parameters that capture the essential ingredients of pair fluctuations across the phase transition.  Most of the published work (see e.g.\ Refs.~\cite{franz1998phase,kwon1999effect,Eckl:2002pair,mayr2005phase,valdez2006single,Chubukov:2007,Berg:2007evolution,valdez2008effects,alvarez2008fermi,Atkinson:2012,Hayward:2014angular}) takes their ensembles from classical Monte Carlo simulations of a GL free energy or two-dimensional XY model. The TDGL calculations are closely related to classical Monte Carlo approaches in that they both generate pair fluctuations that are consistent with the equipartition theorem and they both generate vortex-antivortex excitations near $T_c$.  The explicit time in the TDGL calculations is important for conductivity calculations since it determines the frequency-dependent response to an applied EM field, but is not relevant for the density of states since tunneling is an essentially instantaneous process.  For our purposes, there is no difference between the real TDGL time and fictitious ``Monte Carlo time.''  The two approaches contain the same physics, and we chose the TDGL approach for computational convenience.

Superconducting fluctuations contain a longitudinal component that does not satisfy
the time-independent equation $\nabla \cdot {\bf J} = 0$,
but should satisfy the full continuity equation $\partial_t\rho +\nabla \cdot {\bf J}=0$.
In all of the phenomenological approaches mentioned above, the microscopic BdG Hamiltonian violates the continuity equation for any given order-parameter snapshot, but satisfies it in the ensemble average.
It is  widely believed that this phenomenological approach accurately captures the essential qualitative physics of the superconducting spectrum.  Indeed, the physics is intuitive:  currents generated by phase fluctuations shift the quasiparticle excitation spectrum up or down in energy (so-called Doppler shifts), which broadens features in the density of states and produces gap-filling. (See Franz and Millis \cite{franz1998phase} for a discussion of this physics.)

\begin{figure*} 
    \includegraphics[width=\textwidth]{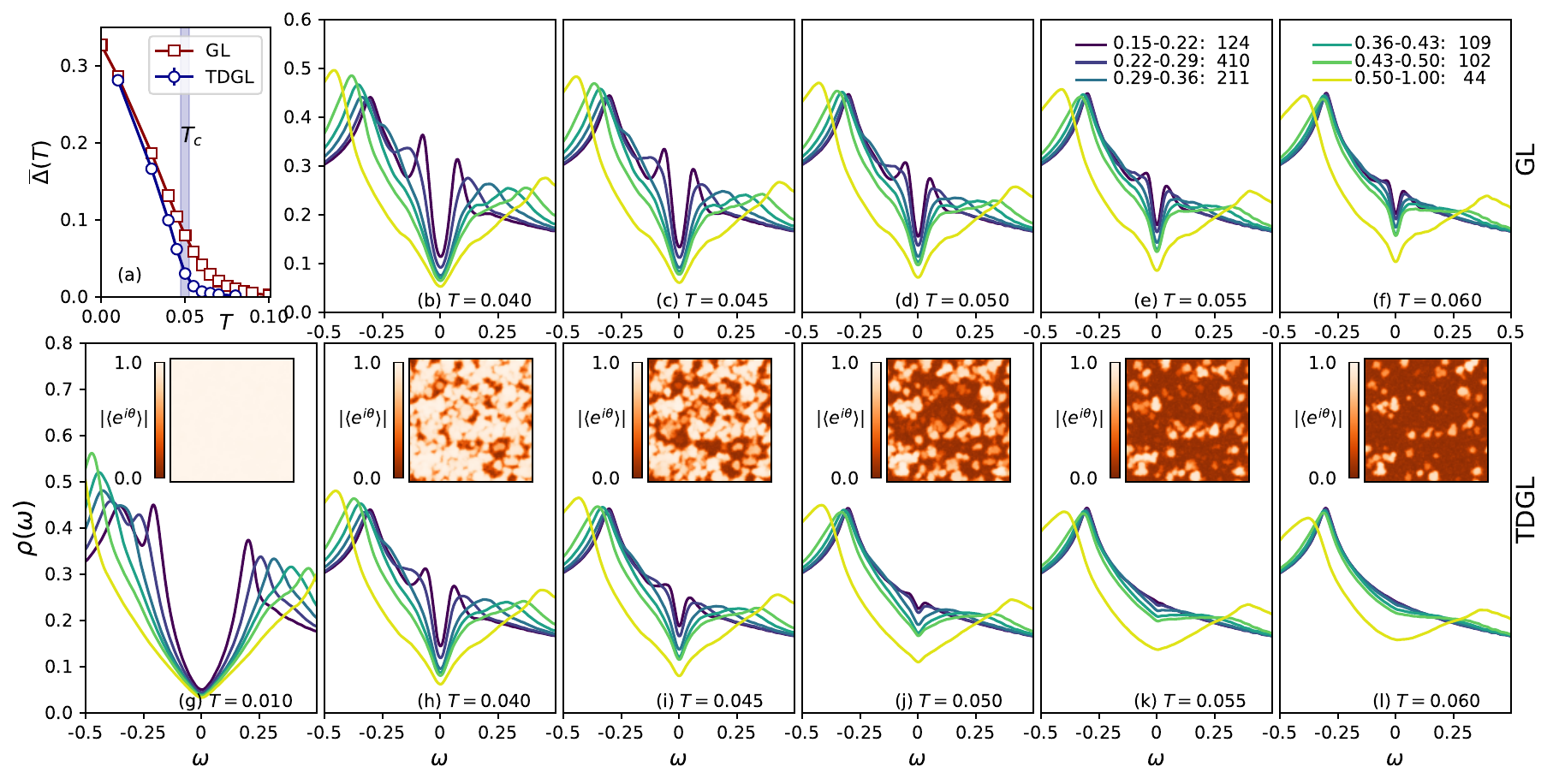}
    \caption{Effect of thermal fluctuations on the density of states. Temperature dependence of (a) the root-mean-square correlation function, (b)-(f) the GL local density of states, and (g)-(l) the TDGL local density of states.  Insets show the time-averaged phase.  In (a), $T_c$ is determined from the scaling of the spatial correlation function (see Section \ref{sec_length_scale}).  The legend shows the spectral-gap ranges for each bin at $T=0.010$ and the number of sites within each bin.}
    \label{fig:LGdos}
\end{figure*}
To characterize the effects of  fluctuations, we show in Fig.~\ref{fig:LGdos}(a) the equal-time pair correlation function,
\begin{equation}
    \overline \Delta(T) = \frac 1N \left[  \sum_{i,j}\langle \Delta_{OP,i}(t) \Delta_{OP,j}(t)^\ast \rangle_t \right ]^{1/2},
\end{equation}
where $\langle \ldots \rangle_{t}$ indicates a time-average. For the (static) GL calculations, $\overline \Delta(T)$ is the spatially-averaged order parameter. 
In Fig.~\ref{fig:LGdos}(a), an extrapolation of the low-$T$ slope would suggest a GL transition near $T=0.06t$; however, $\overline \Delta(T)$ develops upwards curvature near this temperature (a result of the pairing inhomogeneity; similar behavior is seen in BdG calculations\;\cite{Pal:2023}), and only vanishes at $T_\mathrm{GL} \approx 0.14t$. For the TDGL calculations, $\overline \Delta(T)$ is expected to scale as $L^{-\eta(T)}$ (with $\eta(T)$ a $T$-dependent power and $L$ the linear size of the system) below $T_c$ and vanish exponentially above; this provides a simple way to estimate the superconducting transition in finite systems.  In Fig.~\ref{fig:LGdos}(a) we estimate that fluctuations push the transition temperature down by a factor of nearly 3 to $T_c\approx 0.05t$. This value is supported by analysis of the the spatial phase correlations (see Fig.~\ref{fig:SI-corr}).

The remaining panels in Fig.~\ref{fig:LGdos} show the binned LDOS taken from 1000 randomly selected sites, following the same binning scheme as before.  
At low $T$, the GL spectra are indistinguishable from the TDGL spectra in Fig.~\ref{fig:LGdos}(g); %nearly 
$\sim 50\%$ of the sites belong to the 1st or 2nd bin and have a typical $d$-wave spectrum; these are characterized as ``small-gap''. The spectra of the remaining ``large-gap'' sites have a two-gap structure, with a spatially uniform subgap and an inhomogeneous spectral gap.  The subgap and spectral gap are less distinct here than in the BdG calculations because the GL parameters correspond to a less inhomogeneous system.  As $T$ is increased  [Fig.~\ref{fig:LGdos}(b)-(f)], the subgap closes, approximately following the temperature dependence of $\overline \Delta(T)$.

Spectra from the TDGL calculations are shown in Fig.~\ref{fig:LGdos}(g)-(l). The magnitude of the time-averaged phase, $|\langle e^{i\theta_i(t)}\rangle_t|$, is shown in the insets:  it is one (zero) for sites whose correlation time is much longer (shorter) than the simulation time.   Thermal fluctuations are increasingly prominent as $T$ increases, and the phase correlations vanish in progressively larger regions of space with increasing $T$.  Slightly below $T_c$, much of the system is phase-coherent but with patches of phase incoherent pairing, while slightly above $T_c$ the pairing is incoherent everywhere except for small islands of phase coherent superconductivity.    The effect of these fluctuations is to cause a $T$-dependent filling of the subgap; interestingly, while the binned spectra for the large-gap sites are smoothed by fluctuations, the size of the spectral gap changes very little with increasing temperature.  At the highest temperature shown [Fig.~\ref{fig:LGdos}(i)],
there is no trace of superconductivity on  $\sim 75\%$ of the sites in the system. Conversely, the sites on which  spectra have large gaps, which comprise $\sim 25\%$ of our system, have only partially filled gaps:  the proximity-induced subgaps fill quickly with increasing temperature while the high-energy spectral gap persists as a pseudogap [Fig.~\ref{fig:LGdos}(j)-(l)].

\section{Discussion}
As mentioned above, we attribute the subgap to proximity coupling between the large-gap region where the spectrum is measured, and neighboring regions with smaller gaps (see also\;\cite{Sulangi:2025}).  Two-gap structures like those obtained here have been observed repeatedly at low temperatures  in Pb- and La-doped Bi-2201\;\cite{Boyer:2007STM,He:2014STM,Li:2021CDW,Ye:2024Emergent} and Bi-2212\;\cite{Alldredge:2008evolution,Lee:2009STM,Pushp:2009STM}.  In those experiments, the  subgap was interpreted as a superconducting gap and the larger gap was attributed to a non-superconducting pseudogap.  However, the STM data shows that there are small-gap regions whose coherence peaks coincide with the shoulders observed in large-gap regions\;\cite{Ye:2024Emergent}, which is a key requirement for proximity-induced subgaps.  Our calculations predict that these subgaps will fill in when thermal fluctuations wipe out the superconductivity in small-gap regions; that is, the same proximity physics that induces subgaps at low $T$ also wipes them out at high $T$. Conversely, the high-energy coherence peaks, that have been taken for ``pseudogaps", should remain above $T_c$, reflecting a robust and phase-coherent local superconducting patch.  Unfortunately, detailed measurements of the $T$-dependent STM spectra  binned according to gap size have not been reported for the overdoped materials.

We remark that two-gap structures are also ubiquitous in underdoped Bi-2212\;\cite{Schmidt:2011electronic,Pushp:2009STM}  and Bi-2201\;\cite{He:2014STM,Tromp:2023Puddle}.
Superficially, the measured STS spectra look very much like the results presented here. However, these are complicated by the presence of a pseudogap, charge order, and possibly pair density waves\;\cite{Hamidian2016}.
In particular, it is likely that short-range nematic \cite{Lee:2016nematicity}, charge-\;\cite{Atkinson:2018NMR} or pair-density \cite{Choubey:2020atomic,Wang:2021scattering} order contribute to the LDOS.  However, the mechanism is not understood and simple models of intertwined order generally do not reproduce the measured two-gap spectrum\;\cite{Lee:2016nematicity}  (see however \;\cite{Choubey:2020atomic,Wang:2021scattering}); we suggest this points to the importance of pairing inhomogeneity in underdoped cuprates.  Nonsuperconducting order disappears rapidly with overdoping Bi-2201\;\cite{Li:2021CDW} and Bi-2212\;\cite{Fujita:2014simultaneous,Mukhopadhyay:2019evidence}, where our model applies without modification.

We have compared in detail with Bi-2212 and Bi-2201, since
high-quality STM and ARPES data 
is readily available for these systems, and because the distribution of local pairing interactions needed to reproduce gap maps
was known from previous theoretical work \cite{Nunner:2005dopant,melikyan2006gap}  However, we expect our conclusions continue to hold qualitatively for other disordered cuprates, such as HgBa$_2$CuO$_{4+\delta}$, LSCO, and Tl$_2$Ba$_2$CuO$_{6+\delta}$, since the main spectral features identified here are robust against changes in model parameters (see Fig.~\ref{fig:cmpr-model}).
On the other hand, stoichiometric YBa$_2$Cu$_3$O$_7$, which is slightly overdoped, 
is both homogeneous 
 and more three-dimensional, and is not described by our model.

One of the remarkable implications of this work is that the pairing interaction is both local and varies by an order of magnitude over nm length scales.  This can occur, for example, if the pairing bosons are spin fluctuations and the system is near a spin-density wave transition; in that case, inhomogeneity can create large local differences in the dynamical spin susceptibility\;\cite{Romer:2012}.

It was recently shown that a spatially random interaction can produce a ``strange-metal'' linear-in-$T$ resistivity \;\cite{Bashan:2024,patel2024localization,Bashan2025:Extended} like that measured in optimally and overdoped cuprates. Our work suggests that this same strange-metal physics is responsible for anomalous features in the superconducting spectrum.
 
To the extent that our model describes them, overdoped cuprates should exhibit aspects of both BKT and granular superconductivity. The superconducting transition in our model is essentially BKT-like, but with an inhomogeneous distribution of  local phase-disordering temperatures\cite{Benfatto:2009} and gaps\cite{Andersen06} that broadens the transition.  Our use of the Nunner model\;\cite{Nunner:2005dopant} for pair inhomogeneity leads to a dilute concentration of nanoscale regions with high mean-field transition temperatures. (The importance of pair-breaking disorder for the transition in overdoped cuprates is debated \;\cite{Andersen06,Bozovic:2016Dependence,Mahmood:2019microwave,Pal:2023,Lee-Hone:2017disorder,Juskus:2024insensitivity}.) In layered materials, the BKT transition is further broadened by interlayer coupling\;\cite{Benfatto:2007kosterlitz}.

Thus, while superconductivity disappears via a smeared BKT transition, a weak diamagnetic signal will persist well above $T_c$ due to these nanoscale regions.  This provides a natural resolution to an apparent dichotomy between  magnetoresistance\;\cite{Rourke:2011phase} and magnetic susceptibility\;\cite{Sonier:2008,Li:2021Granular} measurements.  The magnetoresistance measurements, which probe the entire sample, will depend on diamagnetic fluctuations in \textit{typical} superconducting regions, while the onset temperature for a diamagnetic response in susceptibility measurements will depend on the GL temperature of the \textit{large-gap} regions.    We therefore believe that  the anomalous spectral properties of the overdoped phase, as measured on inhomogeneous families of cuprates, can be explained by our analysis.

\section{Acknowledgments}
M.A.S. is grateful to the late J. Zaanen for numerous inspiring discussions and collaborations on related projects, and to UFIT Research Computing for providing computational resources and support that have contributed to the research results reported in this publication.
W.A.A. acknowledges the support of the Natural Sciences and Engineering Research Council of Canada (NSERC).  This work was made possible by the facilities of the Shared Hierarchical Academic Research Computing Network (www.sharcnet.ca) and the Digital Research Alliance of Canada.
P.J.H. was supported by NSF-DMR-2231821. A.K. acknowledges support by the Danish National Committee for Research Infrastructure (NUFI) through the ESS-Lighthouse Q-MAT.

\appendix

\section{Bogoliubov-de Gennes Calculations}
%We model the overdoped cuprate by a square-lattice tight-binding Hamiltonian with $d$-wave pairing treated within mean-field theory:
%\begin{equation}
%H = \sum_{ij\sigma} t_{ij}c_{i\sigma}^{\dagger}c_{j\sigma} + \sum_{ij}\Delta_{ij}^{\ast}c_{i \uparrow}c_{j \downarrow} + \text{h.c.}
%\label{eq:hamiltonian}
%\end{equation}
%where $t_{ij} = t \delta_{\langle i,j \rangle} + t' \delta_{\langle\langle i,j\rangle\rangle} - \mu \delta_{i,j}$ with $\delta_{\langle i,j \rangle}$ and $\delta_{\langle\langle i,j\rangle\rangle}$ equal to one for nearest- and next-nearest neighbors, respectively, and zero otherwise.
%The nearest-neighbor hopping, $t = -1$, sets the energy scale, $t' = 0.35$, and $\mu$ is adjusted to maintain an electron density $n=0.85$ per unit cell.  Note that the value of $n$ does not matter for this work so long as the Fermi surface is qualitatively reasonable.
%The order parameter $\Delta_{ij}$ is determined self-consistently from
%\begin{equation}
%\Delta_{ij} = V_{ij}\langle c_{i \uparrow} c_{j \downarrow} \rangle,
%\label{eq:opsc}
%\end{equation}
%where $i$ and $j$ are nearest-neighbor sites.  Self-consistent calculations are performed for $L \times L$ lattices, with $L=40$ with periodic boundary conditions.

\subsection{Pairing Inhomogeneity}
\label{sec_inhomogeneity}
The pairing interaction is assumed to be inhomogeneous because of local modulations due to out-of-plane dopant atoms \cite{Nunner:2005dopant}. The pairing interaction $V_{ij}$ along each nearest-neighbor bond is obtained from the ansatz $V_{ij} = \frac 12 ( V_i+V_j )$, where
\begin{equation}
V_i = V_0 + \sum^{N_i}_{j = 1} \pm V_I \frac{e^{-\frac{\sqrt{|\mathbf{r}_i - \mathbf{r}_j|^2 + r^2_z}}{\lambda}}}{\sqrt{|\mathbf{r}_i - \mathbf{r}_j|^2 + r^2_z}},
\label{eq:yukawa}
\end{equation}
is defined for each lattice site, $N_i$ is the number of impurities, the sign ($\pm$) is selected randomly for each impurity, $V_0 = 0.8$, $V_I = 3.125$, $r_z = 0.7$, and $\lambda = 0.7$.  For these calculations, we take $N_i = 0.075 L^2$.
The created impurity realizations for these parameters are available in the Supplementary Information and are used for impurity averaging.

\subsection{Density of States}
The LDOS at site $i$ is obtained from eigenvalues $E_n$ and eigenvectors $U_{in}$ of the BdG Hamiltonian
\begin{equation}
\rho_i(E) = \frac{1}{L^2 N_k} \sum_{n} \sum_{\bk} |U_{i,n}(\bk)|^2 \delta(\omega - E_{n\bk}),
\label{eq:ldos}
\end{equation}
where we sum over $N_k = 15\times 15$ supercells to minimize finite-size effects \cite{Pal:2023}.
In Eq.~(\ref{eq:ldos}), the $\delta$-function is a Lorentzian of width $\eta = 0.005$.

\begin{table}[ht]
\begin{tabular}{c|c}
     Parameter & Value   \\
     \hline
     $a_0$ &  3  \\
     $T_0$ & 0.01 \\
     $\Delta_0$ & 0.04  \\
     $d$ & 3    \\
      $\Gamma dt$ & $10^{-3}$  \\
     \end{tabular}
     \caption{TDGL Parameters.  All parameters are written in units of the nearest-neighbour hopping $t$, which in cuprates is $\sim 150$~meV.  These parameters correspond to a filling $n=0.85$ and a low-$T$ correlation length $\xi$ of approximately 3 unit cells. }
     \label{table:1}
\end{table}
\section{Time-Dependent Landau-Ginzburg Calculations}
\subsection{Order parameter simulations}
We define singlet bond order parameters $\eta_{i\alpha}$, where $i$ is a site label and $\alpha=x,y$ indicates bonds connected to $i$ in the positive $x$ or $y$ directions. These are related to the usual nearest-neighbour bond order parameters by
\begin{eqnarray}
\Delta_{i+\alpha, i} = \Delta_{i, i+\alpha} &=& V\langle c_{i+\alpha\downarrow} c_{i\uparrow}\rangle = \eta_{i\alpha}, \label{eq:Dij}
\end{eqnarray}
where $i+\alpha$ indicates the nearest-neighbour site to $i$ along the direction $\alpha$, and $V$ is a mean-field pairing interaction.  Rather than solving the mean-field equations directly, which is computationally prohibitive for large system sizes, we obtain the superconducting order parameter from solutions of the TDGL equations with thermal noise.  That is, we write
\begin{eqnarray}
\frac{\partial \eta_{i\alpha}}{\partial t} &=& -\Gamma \frac{\delta {\cal F}}{\delta \eta_{i\alpha}^\ast} + \sqrt{\frac{k_BT\Gamma}{dt}} \, \zeta_i \label{eq:TDGL1}
\end{eqnarray}
where ${\cal F}$ is the GL free energy, $\Gamma$ is a relaxation rate, and $\zeta_i$ is a site-dependent gaussian-distributed complex random variable, taken from the distribution
\begin{equation}
    P(\zeta_i) = \frac{1}{2\pi} e^{-|\zeta_i|^2/2},
\end{equation}
and satisfying $\langle \zeta_i(t)^\ast \zeta_j(t')\rangle = 2\delta_{i,j}\delta_{t,t'}$.  Here, $t$ and $t'$ are discrete time variables with time step $dt$, and the time derivatives are understood as finite differences.  The specific value of $\Gamma$ does not matter for the current calculations so long as the fractional changes in $\eta_{i\alpha}$ at each time step are small.  In the absence of thermal noise ($\zeta_i=0$), and at large times, Eq.~(\ref{eq:TDGL1}) asymptotically approaches the usual  GL equation,
\begin{eqnarray}
    \frac{\delta {\cal F}}{\delta \eta_{i\alpha}^\ast} = 0.
    \label{eq:GL}
\end{eqnarray}

The free energy  is
\begin{widetext}
\begin{eqnarray}
{\cal F} &=& a_0 \sum_i \sum_{\alpha=x,y} \left [  \left ( \frac{T-T^\ast_{i\alpha}}{T_0} \right ) |\eta_{i\alpha}|^2
+ \frac 1{2\Delta_0^2}|\eta_{i\alpha}|^4 +  \frac{\xi_d^2}{2} |\nabla \eta_{i\alpha}|^2 \right ] \nonumber \\
 && + \frac{d}{4} \sum_i  \big [ 4(|\eta_{ix}|^2 + |\eta_{iy}|^2) +  ( \eta_{ix} + \eta_{i-x, x} )^\ast (\eta_{iy} + \eta_{i-y, y}) + ( \eta_{ix} + \eta_{i-x, x} )(\eta_{iy} + \eta_{i-y, y})^\ast  \big ]
\label{eq:S}
\end{eqnarray}
\end{widetext}
In this expression, $a_0$, $d$, $\Delta_0$, and $\xi_d$ are GL parameters, $T_0$ is a characteristic temperature scale for pairing, and $T_{i\alpha}^\ast$ is the local GL temperature below which the singlet pairing interaction along the bond $(i, i+\alpha)$ becomes attractive.
Note, however, that $T_{i\alpha}^\ast$ is not the local critical temperature, even at the mean-field level, because the order parameter on the bond $(i,i+\alpha)$ depends on the pairing interaction in the neighborhood of the bond.  The parameter $d$ determines whether the dominant pairing channel has $d$-wave ($d>0$) or extended $s$-wave ($d<0$) symmetry.  In the former case, the lowest-energy GL solution for a homogeneous pairing interaction ($T_{i\alpha}^\ast=T_0$) has $d$-wave symmetry with $\eta_x = -\eta_y = \Delta_0$ at $T=0$.  An extended $s$-wave solution is also possible, with $\eta_x = \eta_y = \Delta_0\sqrt{(a_0-2d)/a_0}$, however this has higher energy.   Values for the GL parameters are shown in Table~\ref{table:1}. Both the temperature scale $T_0$ and the pairing scale $\Delta_0$ are somewhat large for a quantitative description of overdoped cuprates; however, these choices help mitigate finite size effects in our simulations.

The first step in the TDGL simulations is to find the solutions to the GL equations, Eq.~(\ref{eq:GL}), by iterating Eq.~(\ref{eq:TDGL1}) to self-consistency with the thermal noise term set to zero.  The thermal noise is then turned on, and a typical TDGL simulation runs for $10^5$ time steps.  Thermal equilibrium is typically achieved within a few hundred time steps.  The density of states can be obtained for both the GL and  TDGL simulations; in the latter case, the density of states is averaged over 100 snapshots taken from the simulation.

\subsection{Density of States}
To obtain the local density of states for a given order-parameter snapshot, we write a mean-field Hamiltonian,
\begin{equation}
{\bf H} = \left [ \begin{array}{cccc}
{\bf H}_{11} & {\bf H}_{12} & \ldots \\
{\bf H}_{21} & {\bf H}_{22} & \\
\vdots & & \ddots \\
&&& {\bf H}_{NN}
\end{array}\right ],
\end{equation}
where ${\bf H}_{ij}$ is a  $2\times 2$ block corresponding to a particular pair of sites $i$ and $j$, with
\begin{equation}
{\bf H}_{ij} = \left [ \begin{array}{cc} t_{ij} & \Delta_{ij} \\ \Delta^\ast_{ji} & -t_{ji} \end{array} \right ].
\label{eq:hij}
\end{equation}
In Eq.~(\ref{eq:hij}), the hopping matrix elements $t_{ij}$ are the same as for the BdG calculations.
The superconducting order parameters $\Delta_{ij}$ are only nonzero for nearest-neighbor sites and are either taken from solutions of the GL equations or from snapshots of the TDGL equations, using Eq.~(\ref{eq:Dij}).  To mitigate finite-size effects, results in this work are shown for a $300\times 300$ lattice.  This precludes standard matrix diagonalization approaches to obtain the density of states.  Rather, we use Haydock's recursion algorithm to find the density of states at selected individual sites \cite{Haydock:1994}.  For a single configuration $\Delta_{ij}$, the recursion algorithm evaluates the local density of states at site $i$,
\begin{equation}
\rho_i(\omega) = -\frac{1}{\pi}\mbox{Im } [\omega + i\delta - {\bf H}]^{-1}_{ii}
\label{eq:dos}
\end{equation}
where $[\ldots]^{-1}$ denotes a matrix inverse and $\delta=0.02$ is a Lorentzian broadening parameter.

Inhomogeneity is achieved through spatial variations of the coefficient $T^\ast_{i\alpha}$ of the quadratic term in the GL energy, Eq.~(\ref{eq:S}), with
\begin{align}
    T_{i\alpha}^* &= (T_i^* + T_{i+\alpha}^*)/2, \label{eq:Ti*_alpha}\\
    T_{i}^* &= T_0 + \delta T^*\sum_sf_{is}, \label{eq:Ti*}
\end{align}
where the sum is over impurities, $f_{is} = \exp(-r_{is}/\kappa)/r_{is}$  for impurity $s$, $r_{is}$ is the distance between the impurity and site $i$, $\kappa$ is a characteristic length scale over which pairing is enhanced, and $\delta T^\ast$ sets the scale for the local enhancement. We set $\delta T^\ast = 10T_0$ and $\kappa = 2.8$ lattice constants, take the dopant atoms to sit a distance $0.8$ lattice constants above the plane, and set the dopant concentration to be $2.5\%$. The resulting distribution of $T_{i\alpha}^*$ used in the simulations is available in the supplementary information.

\subsection{BKT scaling}
\begin{figure}[tb]
    \centering
    \includegraphics[width=\columnwidth]{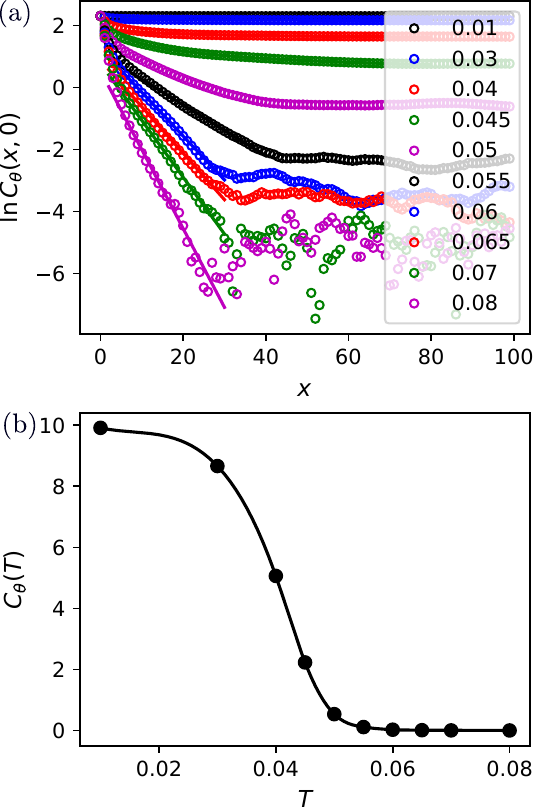}
    \caption{(a) Pair-phase correlation  function $C_\theta(\br)$ along $\br=(x,0)$ at different temperatures.  Symbols show numerical results for Eq.~(\ref{eq:Ctheta}) taken from TDGL data for cuts along the $x$-axis, while solid curves are fits to the asymptotic forms in Eq.~(\ref{eq:BKT}).   Fits are to the power-law form for $T\leq 0.045$ and the exponential form for $T\geq 0.55$.  The data at $T=0.050$ cannot be fitted by either form. (b) Temperature dependence of the spatially averaged phase correlation function.  The average is performed over sites separated by $50 < |\br| < 150$ lattice constants. }
    \label{fig:SI-corr}
\end{figure}
In homogeneous systems, the BKT transition is characterized by a power-law pair correlation function below $T_\mathrm{BKT}$ and an exponential correlation function above $T_\mathrm{BKT}$.  We define the phase of the $d$-wave component of the order parameter via
\begin{equation}
    e^{i\theta_j(t)} = \frac{\Delta_{d,j}(t)}{|\Delta_{d,j}(t)|}.
\end{equation}
The phase correlation function is then
\begin{equation}
    C_{\theta}(\br) = \frac 1N \sum_{j} \left \langle e^{i[\theta_j(t) -\theta_{j+\br}(t)]} \right \rangle
    \label{eq:Ctheta}
\end{equation}
where $\langle \ldots \rangle$ denotes a time-average.  Figure~\ref{fig:SI-corr} shows $C_\theta(\br)$ for $\br = (x,0)$ (i.e. for a cut along the $x$-axis).  The correlation function is fitted to expressions for the homogeneous system, namely
\begin{equation}
    C_\theta(\br) = \left \{ \begin{array}{ll}
    a \left [ x^{-\eta(T)} + (L-x)^{-\eta(T)} \right], & T<T_\mathrm{BKT} \\
    a' e^{-x/\xi(T)}, & T>T_\mathrm{BKT}
    \end{array} \right .
    \label{eq:BKT}
\end{equation}
with $a$, $\eta$, $a'$, and $\xi$ taken as fitting parameters.  These fits are shown as solid lines in Fig.~\ref{fig:SI-corr}(a).  Convincing  fits are obtained for the power law when  $T\leq 0.045$ and for the exponential when $T \geq 0.055$; however the data at $T=0.050$ is not well-reproduced by either asymptotic form, indicating that the superconducting transition temperature is $T_c \sim 0.050$.

As further confirmation, we plot the temperature dependence of $C_\theta(\br)$, averaged over separations $50<|\br|<150$ in Fig.~\ref{fig:SI-corr}(b).  This figure shows that phase stiffness is lost near $T=0.055$.

\section{Sensitivity to Parameters}
\begin{figure}[tb]
    \includegraphics[width=\columnwidth]{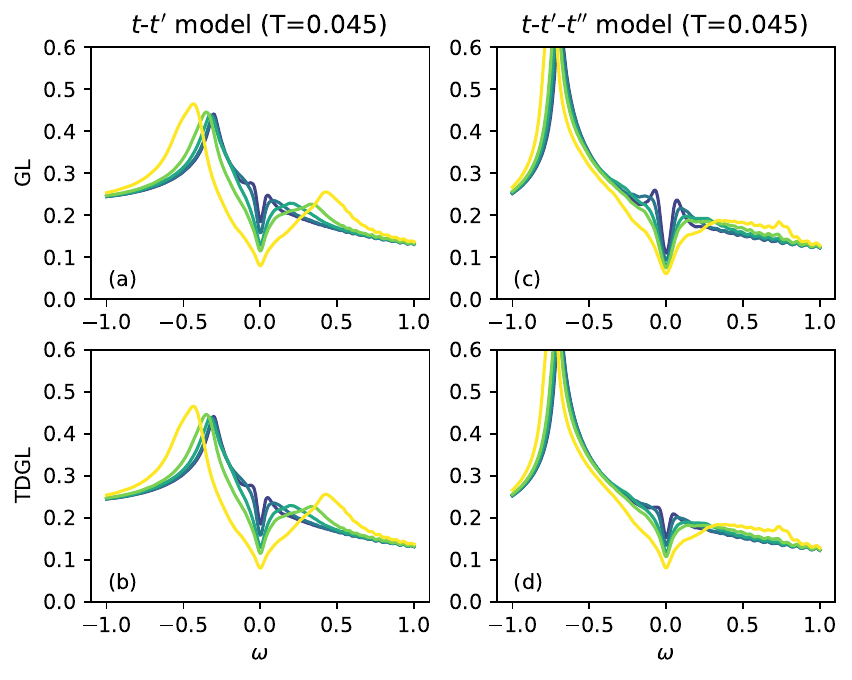}
    \caption{Density of states at $T=0.045$ for (a), (b) the $t$-$t'$ model shown in the main text, and (c), (d) a $t$-$t'$-$t''$ model.  Results are shown for (a), (c) static GL calculations and (b), (d) thermally fluctuating TDGL calculations.  The calculations use the same inhomogeneous GL parameters, $T_{i\alpha}^\ast$, as in the main text but different band structures for the calculated LDOS.  The LDOS is binned such that the same sites are allocated to each bin for the two models.}
    \label{fig:cmpr-model}
\end{figure}

\begin{figure}[tb]
	\includegraphics[width=\columnwidth]{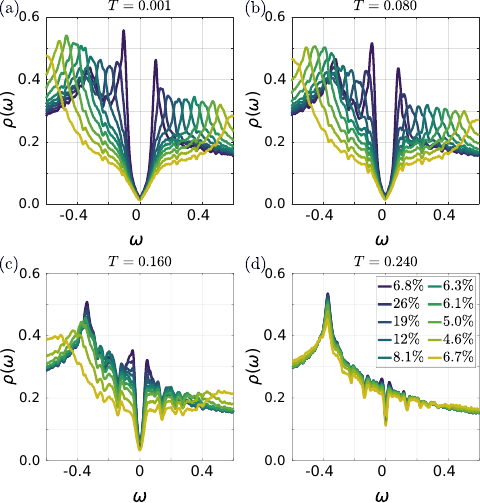}
	\caption{Bin-averaged LDOS (bottom row) with inhomogeneity parameters $V_0 = 0.8$, $V_I = 3.125$, $r_z = 1.4$, and $\lambda = 1.4$. 50 random pairing realizations and $15\times 15$ supercells are used.  As with the inhomogeneity model shown in the main text, the sites are binned by the value of $\Delta_{SG}$ at $T=0.001$, and the LDOS within each bin is averaged over.  The same sites are in each bin at all temperatures. Ten equal-width bins spanning $0.054 < \Delta_{SG} < 0.605$ are used; the legend shows the fraction of sites in each bin and darker colors correspond to lower-energy bins.}
	\label{fig:cmpr-model-gapresolvedldos}
\end{figure}

\begin{figure}[tb]
	\includegraphics[width=\columnwidth]{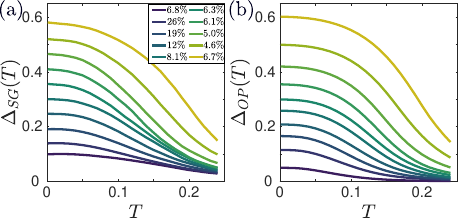}
	\caption{The binned spectral gap (a) and order parameter (b) as functions of temperature for the inhomogeneity model with $N_i = 0.075L^2$, $V_0 = 0.8$, $V_I = 3.125$, $r_z = 1.4$, and $\lambda = 1.4$. At the lowest measured temperature, each site is sorted according to its local value of $\Delta_{SG}$ and put into one of ten bins; the values of $\Delta_{SG}$ and $\Delta_{OP}$ are then averaged over each bin at each temperature.}
	\label{fig:cmpr-model-binned-sg-op}
\end{figure}

\subsection{Band-Structure Parameters}

Figure~\ref{fig:cmpr-model} compares the binned LDOS for the band structure used in the main text (the $t$-$t'$ model) and another common tight-binding model for the cuprates (the $t$-$t'$-$t''$ model).  For the latter model, we take $t=-1$, $t'=0.4$, and $t''=-0.05$.  The main difference between the two models is that the van Hove singularity lies further from the Fermi level in the $t$-$t'$-$t''$ model, although their electron densities are the same ($n=0.85$ electrons per unit cell).  Since the primary effect of doping in our model is to move the van Hove singularity towards or away from the Fermi level, these two models also provide insight into the effects of doping.

Although there are quantitative differences between the two GL calculations, Fig.~\ref{fig:cmpr-model}(a) and (c), the important qualitative structures are the same:  both exhibit a two-gap structure with a homogeneous subgap and an inhomogeneous spectral gap.  Similarly, the TDGL calculations show identical trends, namely that the subgap fills relative to the GL gap but that the high-energy spectral gap is largely unaffected by thermal fluctuations.

\subsection{Pairing Interaction Length Scale}
\label{sec_length_scale}
Figure~\ref{fig:cmpr-model-gapresolvedldos} shows the binned LDOS for a modified version of the inhomogeneity model used in the main text with a larger characteristic length scale. Here, we set $0.075L^2$, $V_0 = 0.8$, $V_I = 3.125$, $r_z = 1.4$, and $\lambda = 1.4$ in Eq.~\ref{eq:yukawa}. Notice that only $r_z$ and $\lambda$ are changed from the values used in the main text (both are doubled in this case); all other parameters are kept the same. It can be seen that much of the same striking behavior seen in the main text (i.e., the two-gap structure consisting of low-energy homogeneity and shoulders and an inhomogeneous spectral gap, and the overall anticorrelation between the spectral gap and the coherence peak) remains present even when the inhomogeneity length scale is doubled.

We show in Figure~\ref{fig:cmpr-model-binned-sg-op} the $T$-dependence of $\Delta_{SG}$ and $\Delta_{OP}$, both of which have been binned using the same procedure as in the main text. Much of what is observed in the main text is also seen here: at high temperatures, $\Delta_{SG}$ is homogeneous, whereas $\Delta_{OP}$ continues to show large variations.

Some differences between this model and the one used in the main text are: a) the spectral-gap-coherence-peak height anticorrelation is less strong in the long-ranged inhomogeneity model compared to the short-ranged one; and b) the gap persists to a larger temperature in the long-ranged model compared to the short-ranged one. However, the important qualitative features we have highlighted in the main text remain robust to changes in the inhomogeneity length scale. We defer an in-depth discussion of the length-scale dependence of the spectra and the superfluid density to a future manuscript.

\bibliography{InhomSC}
\end{document}